\documentclass[%
 reprint,
 amsmath,amssymb,
 aps,
]{revtex4-2}

\usepackage[colorlinks=true,linkcolor=blue,citecolor=blue,urlcolor=blue]{hyperref}

\usepackage{graphicx}
\usepackage{dcolumn}
\usepackage{bm}
\usepackage[mathlines]{lineno}

\usepackage{soul}
\usepackage{yfonts}
\usepackage{amsmath}
\usepackage{xcolor}
\usepackage{float}
\usepackage{amsfonts}
\usepackage{amsthm}

\begin{document}

\preprint{APS/123-QED}

\title{Introducing a novel $Z_{4n}$-detection scheme to enhance the performance of quantum LiDAR systems}

\author{ Priyanka~Sharma$^{1}$}
\author{ Manoj~K~Mishra$^{2}$}
\author{ Devendra~Kumar~Mishra$^1$}
\email{Corresponding author: kndmishra@gmail.com}
\affiliation{$^1$Department of Physics, Banaras Hindu University, Varanasi-221005, India}
\affiliation{$^2$Space Applications Centre, Indian Space Research Organization (ISRO), Ahmedabad, Gujarat, India}

\vspace{10pt}

\date{\today}

\begin{abstract}
In a quantum LiDAR system, to achieve a better resolution and sensitivity, detection scheme plays an important role. We propose a novel detection scheme in which the photo detector considers only the $4n$ number of photons, where $n \in \mathbb{N}$, as a click and the rest of them as a no-click. Similar to the $Z$-detection scheme, where we get a click for any number of photons, we termed this measurement as $Z_{4n}$-detection scheme. By employing superposition of four coherent states (SFCS) and vacuum as input we investigate the performance of Mach-Zehnder interferometer (MZI) based quantum LiDAR systems. We found a significant enhancement in resolution and broader working point for the phase sensitivity in comparison to the $Z$-detection scheme. Our findings highlight the advantages of our approach and suggest promising advancements in the field of quantum LiDAR sensing technology, providing a pathway for more accurate and sensitive measurement capabilities.
\end{abstract}

\maketitle


\textit{Introduction:}\label{section1}
LiDAR (Light Detection and Ranging) is a remote sensing technology which uses laser light to measure distances and to draw a precise, three-dimensional maps of the environment \cite{Collis:70, Weibring:03}, in real time. It is broadly uses in the field of aerospace \cite{atmos12070918}, agriculture \cite{RIVERA2023107737}, defence \cite{Fan_2021_ICCV}, autonomous vehicles \cite{app9194093}, etc. The versatility of LiDAR applications makes it a useful technology and opens up a wide range of research for further improvement.

The fringe resolution obtained by a laser-based LiDAR system is directly proportional to the $\lambda/2$, where $\lambda$ is the wavelength of the laser light, and saturates the classical limits \cite{dowling2009quantum}. In order to improve the resolution and sensitivity, quantum metrological techniques \cite{1976Helstrom, giovannetti2006quantum} are widely used \cite{dowling2009quantum, SHARMA2024131028, Shukla:22, SHUKLA2024100200}. With the emergence of quantum metrology \cite{1976Helstrom, giovannetti2006quantum}, the quantum version of LiDAR has been proposed to achieve better precision and resolution in comparison to classical LiDAR and stated as quantum LiDAR (QLiDAR) \cite{dowling2009quantum, slepyan2021quantum}. Recent advancements have focused on enhancing the performance of QLiDAR systems by exploring various quantum states and detection schemes \cite{glauber1963coherent, gerry2007parity, gerry_knight_2004, 1976Helstrom, Guo_2015}. Quantum states like N00N state \cite{PhysRevLett.85.2733}, entangled Fock states \cite{PhysRevA.78.063828}, Schrodinger's cat states \cite{WANG20163717}, multi-photonic states \cite{Sharma:24}, etc. are improving the QLiDAR resolution and sensitivity. 
On the other hand, detection scheme also plays an important role \cite{mitchell2004super, PhysRevLett.98.223601, Sharma:24, 1wb7-1z1b} in  the improvement. For example, a coherent state-assisted conventional MZI gives the fringe resolution ($\Delta x$) of $\lambda/2$ in intensity measurement \cite{born2013principles, PhysRevLett.85.2733}. While using photon-number-resolving detectors, resolution becomes $N$ fold, i.e., $\Delta x \propto \lambda/2N$ \cite{Gao:10, PhysRevLett.111.033603}, where $N$ is the mean number of photons in the input coherent state. 

Photon-number-resolving (PNR) detection has emerged as a key tool in quantum-enhanced sensing. Its study in theory as well as in the experimental domain has demonstrated its significance \cite{DiGiuseppe2003Pairs, Fitch2003PNR, Munro2005QND, Gerrits2010CSS, Nunn2021Zero, Wein2024PhotonCounting}. For example, Munro \textit{et al.} \cite{Munro2005QND} proposed a high-efficiency ($\sim 99\%$) quantum non-demolition (QND) measurement scheme based on strong Kerr nonlinearities, which precisely counts the photon without destroying the quantum state. Furthermore, realistic models of PNR detectors, including finite quantum efficiency and noise, have been developed to accurately characterize the photocount statistics \cite{Sperling2012Photocounting, Jonsson2019PNR}.

On the quantum state side, the superposition of coherent states has attracted considerable interest due to their strong nonclassical features and experimental feasibility. In article \cite{WANG20163717}, Wang \textit{et al.} showed that with the superposition of two coherent state, even coherent superposition state (ECSS), defined as $(|i\alpha\rangle + |-i\alpha\rangle)$ \cite{DODONOV1974597}, improves the resolution and sensitivity compared to the single coherent state (CS), $|\alpha\rangle$, as an input by using photon counting measurement schemes. Further in our previous work \cite{Sharma:24}, we showed that the superposition of four coherent state (SFCS) gives better resolution and phase sensitivity than the ECSS as the input states. In the resolution analysis, we had applied parity and $Z$-detection schemes and found significant improvement in it. In principle, in parity detection, the detector counts the even and odd numbers of photons, and in $Z$-detection, detector counts the zero and non-zero numbers of photons. 

In this letter, we are proposing a detection scheme, in which detector counts the $4n$ number of photons, where $n \in \mathbb{N}$, and we will term it as the $Z_{4n}$-detection scheme throughout the letter. By implementing the $Z_{4n}$-detection scheme, we study the resolution and phase sensitivity of QLiDAR systems. The idea behind introducing this detection scheme just comes out from the `superposition of four coherent states (SFCS)' \cite{Mishra2021QuquatsAS}. The nonclassical properties of SFCS applicable in quantum technology have been studied in our previous article  \cite{Sharma:24}, and an experimental scheme for the generation of SFCS has been proposed by Mishra \textit{et al.} in \cite{Mishra2021QuquatsAS}.

\begin{figure}
\centering
\includegraphics[width=0.5\textwidth]{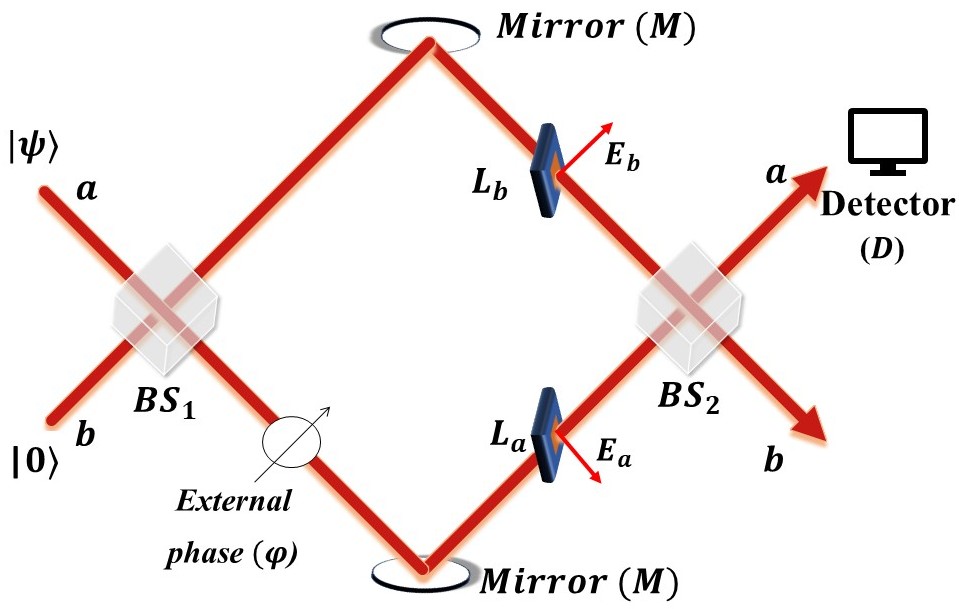}
\caption{\label{fig:1} MZI has two input (modes $a,~b$) and two output ports including two 50:50 beam splitters ($BS_1$ and $BS_2$), two mirrors ($M$), a phase shifter $\phi$, detector $(D)$. The two fictitious beam splitters  ($L_a$ and $L_b$) whose corresponding modes are $E_a$ and $E_b$, respectively, mimic the photon loss inside the two arms of MZI.}
\end{figure}

\textit{Analysis with SFCS and $Z_{4n}$-detection scheme:}\label{section 2} Considering SFCS as an input state, defined as
\begin{equation}
|\Psi\rangle= \mathcal{N}(C_1|\alpha\rangle+C_2|i\alpha\rangle+C_3|-\alpha\rangle+C_4|-i\alpha\rangle),\label{eq2}
\end{equation}
where $|e^{i\theta}\alpha\rangle$ is an eigen state of the annihilation operator with amplitude $|\alpha|$ and phase $\theta$ and $C_i$ ($i=1,2,3,4$) are the real coefficients. $\mathcal{N}$ is the normalisation constant, given as

\begin{equation}
 \begin{split}
 |\mathcal{N}|= \left(X +2e^{-|\alpha|^2}\Big(Y+V\cos|\alpha|^2\Big)\right)^{\frac{-1}{2}},
 \end{split}\label{N0}
\end{equation}
where,
$ X=(|C_1|^2+|C_2|^2+|C_3|^2+|C_4|^2)$, $Y=(|C_1||C_3|+|C_2||C_4|),~V=(|C_1|+|C_3|)(|C_2|+|C_4|)$.\label{eq5}

From Eq. \eqref{eq2}, we can see that ECSS ($C_1=C_3=0$ and $C_2=C_4=1$) and CS ($C_2=C_3=C_4=0$) are the special cases of the SFCS. We study the effect of our proposed detection scheme on a simple case of SFCS by considering equal contributions of the four coherent states, i.e., $C_1=C_2=C_3=C_4=1$. And, to make the study wider, we see the effect of our proposed detection scheme with CS and ECSS states along with SFCS. Since SFCS is a general state and CS and ECSS are the special cases, we perform the calculation for SFCS, and by choosing proper condition, we get the result for the special cases.

Our initial state of the system can be written as \cite{Sharma:24}
\begin{equation}
|\psi_{in}\rangle=|\Psi,0,0,0\rangle_{a,b,E_a,E_b}.
\end{equation}
Here $|\Psi\rangle$ is the SFCS at input mode $a$ and $|0\rangle$ at input mode $b$. Mode $E_a$ and $E_b$ are representing the photon loss as shown in Fig. \ref{fig:1}. By following the state transformations throughout the interferometer discussed in ref. \cite{Sharma:24}, final state of the system at output of MZI reads
\begin{equation}
\begin{split}
     |\psi_{out}\rangle= \mathcal{N}\left(|F_1\rangle+|F_2\rangle + |F_3\rangle +|F_4\rangle\right)_{a,b,E_a,E_b}.\label{20out}
\end{split}
\end{equation}
Here $|F_1\rangle,~|F_2\rangle,~|F_3\rangle,~|F_4\rangle$ are given in Eq. \eqref{eqA3} and $a,~b$ are the output modes (Fig. \ref{fig:1}). Eq. \eqref{20out}, contains the information about the unknown phase change during the probe evolution inside the interferometer and this will be used to calculate the resolution and phase sensitivity of the MZI in this case. In order to study the resolution, full width at half maximum (FWHM) approach is used \cite{Sharma:24}. In this approach, variation with $\phi$ in `mean value of observable (MVO)', w.r.t. output state, is plotted. Where FWHM of the MVO is directly proportional to the resolution of the system (Since, $\Delta x \propto \text{FWHM}$) \cite{Sharma:24}.

In order to measure the MVO, w.r.t. output state, we need to calculate the density operator for the output state of the system. Density operator for the output state can be written as
\begin{equation}
    \hat{\rho} = |\psi_{out}\rangle_{a,b}~ _{a,b}\langle\psi_{out}|.\label{22abc}
\end{equation}
Probability of detecting $l$ photons at the output mode $a$ and $m$ photons at the output mode $b$ can be written as 
\begin{equation}
 P(l,m)=_{a,b}\langle l,m|\hat{\rho}|l,m\rangle_{a,b} \label{eq6}.
\end{equation}
In order to find the probability of getting $l$ photon number at output mode $a$, we sum over $m$ and can be written as 
\begin{equation}
P(l)=\sum_{m}P(l,m).\label{p9}
\end{equation}
Therefore, from Eqs. (\ref{eq6}) and (\ref{p9}), the probability of getting $l$ photon numbers at output mode $a$ is 

\begin{equation}
 \begin{split}
P(l) =|\mathcal{N}|^2e^{-|\alpha|^2}\frac{p^l}{l!} \left(Xe^{-p+|\alpha|^2}\right.\\
\left.+2V \cos\Big(q+\frac{l\pi}{2}\Big)+ 2 Ye^{-q+il\pi}\right).
 \end{split}  \label{eq p18}
\end{equation}
Where
\begin{equation}
    \begin{split}
        q=|\alpha|^2\Big(|t|^2cos^2\left(\frac{\phi}{2}\right)+|r|^2\Big),~p= |\alpha|^2|t|^2sin^2\left(\frac{\phi}{2}\right),\label{eq29}
    \end{split}
\end{equation}
with reflection ($r$) and transmission ($t$) coefficients of the fictitious beam splitters, satisfies $|r|^2 + |t|^2 = 1$.

\textit{$Z_{4n}$-detection:}
Analogs to $Z$- detection scheme, in this measurement scheme, we explore the data obtained by a standard single-photon detector at the output port $a$ that cannot differentiate between different photon numbers \cite{PhysRevA.90.013807, Gao:10}. It has only two outcomes, $4n$ number of photons ($n \in \mathbb{N}$) and any number of photons ($\neq 4n$) with the associated probabilities $P(4n)$ and $P(\neq 4n) = 1 - P(4n)$, respectively. For a better approximation, analogs to the $Z$- detection scheme, we can take the difference between these two probabilities, i.e., $P(4n)-P(\neq 4n)$. This gives us $P(4n)-P(\neq 4n)=2P(4n)-1$, so considering only $P(4n)$ gives the better approximation. The observable for $4n$ photon detection can be defined by the $4n$-photon projection operator $\hat{Z}_{4n}=|$4n$\rangle_{a~a}\langle 4n|$ and its expectation value w.r.t. $|\psi_{out}\rangle_{a,b}$, can be written as
\begin{equation}
     \langle \hat{Z}_{4n}\rangle = P(4n).
\end{equation} 
From Eq. \eqref{eq p18},
\begin{equation}
\begin{split}
\langle \hat{Z}_{4n}\rangle =|\mathcal{N}|^2e^{-|\alpha|^2}\frac{p^{4n}}{4n!} \left(Xe^{-p+|\alpha|^2}\right.\\
\left.+2V \cos\Big(q+\frac{4n\pi}{2}\Big)+ 2 Ye^{-q+i4n\pi}\right).\label{z4nphase}
 \end{split} 
\end{equation}

In order to calculate the expectation value for $Z$-detection scheme \cite{Sharma:24}, put $n=0$ in Eq. \eqref{z4nphase}, we get
\begin{equation}
\begin{split}
\langle \hat{Z}\rangle =|\mathcal{N}|^2e^{-|\alpha|^2} \left(Xe^{-p+|\alpha|^2}\right.\left.+2V \cos\Big(q\Big)+ 2 Ye^{-q}\right).\label{eq 20}
 \end{split} 
\end{equation}
Eq. \eqref{z4nphase} and \eqref{eq 20} are the expectation values of $\hat{Z}_{4n}$ and $\hat{Z}$ w.r.t. $|\psi_{out}\rangle_{a,b}$. In order to perform a wide range study, we examine our proposed detection scheme for CS, ECSS, and SFCS and compare the results with $Z$-detection scheme results. To see the effect of input photon numbers on the resolution and phase sensitivity, we will choose $\Bar{N}=3$ and $\Bar{N}=100$, where $\Bar{N}$ is the input mean photon number can be calculated by using the expectation value expression $\langle\psi_{in}|\hat{a}^\dagger\hat{a}|\psi_{in}\rangle$, where $\hat{a}(\hat{a}^\dagger)$ is annihilation (creation) operator of the input ports.

\begin{figure}[h!]
\centering
\includegraphics[width=.40\textwidth]{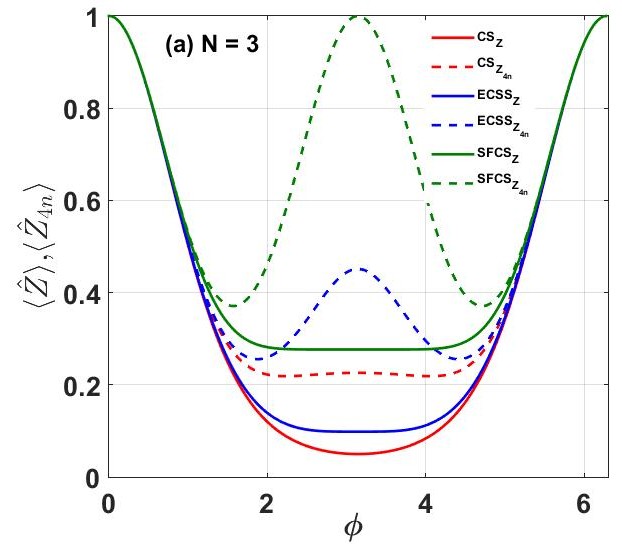}
\includegraphics[width=.40\textwidth]{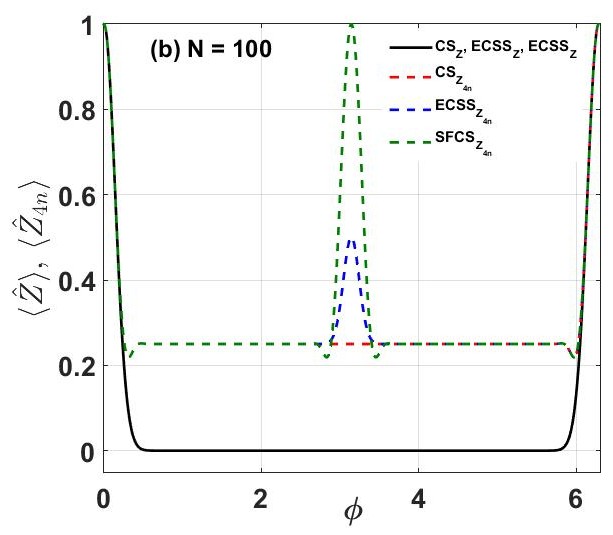}
\caption{\label{fig_4} Plot shows the variation of $\langle Z\rangle$ and  $\langle Z_{4n}\rangle$ with $\phi$ for CS, ECSS and SFCS in lossless ($r = 0$) case. Here we have taken mean photon number (a) $\Bar{N}=3$ and (b) $\Bar{N}=100$.}
\end{figure}

\textit{Discussion:} To study the resolution in both detection schemes, we plot the variation of $\langle\hat{Z}_{4n}\rangle$ and $\langle\hat{Z}\rangle$ with $\phi$ for $\Bar{N} = 3$ and $\Bar{N} = 100$ as shown in Fig. \ref{fig_4}. From Fig. \ref{fig_4}, we can draw the following important results:\\
(i) $Z_{4n}$-detection scheme shows double foldness in compare to the $Z$-detection scheme for SFCS and ECSS. (ii) In case of CS, we get no extra foldness in resolution. (iii) On increasing the mean number of photons, FWHM decreases, which implies that resolution increases. 

\begin{figure}[h!]
\centering
\includegraphics[width=.45\textwidth]{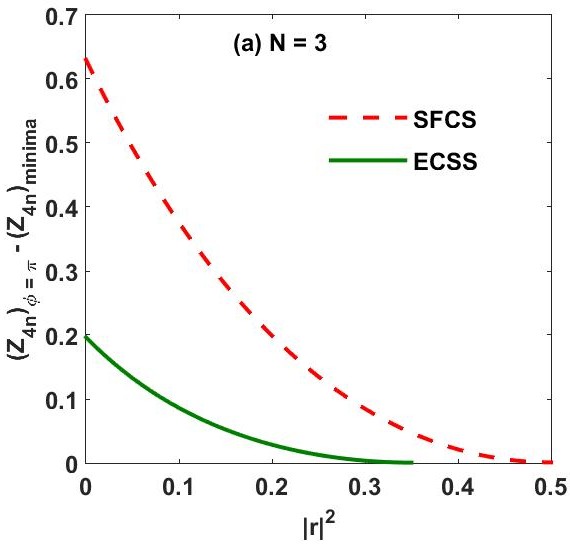}
\includegraphics[width=.45\textwidth]{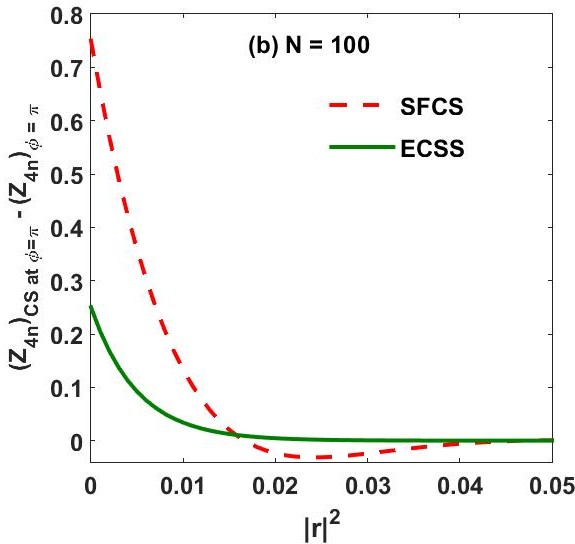}
\caption{\label{fig_5} Plots show the effect of photon loss on the resolution peak for SFCS and ECSS. Here we have taken mean photon number (a) $\Bar{N}=3$ and (b) $\Bar{N}=100$. For $\Bar{N}=3$ case, we consider the difference quantity $(Z_{4n})_{\phi=\pi}$ - $(Z_{4n})_{\text{min}}$ with respect to the loss parameter $|r|^2$. For $\Bar{N}=100$ case, we consider the difference quantity $(Z_{4n})_{CS \ \text{at} \ \phi=\pi}$ - $(Z_{4n})_{\phi=\pi}$ with photon loss $|r|^2$.}
\end{figure}

The results in Fig. \ref{fig_4} show the photon lossless case. In the presence of photon loss, the resolution peak starts to decrease with increase in $r$ and becomes invisible for higher photon loss. To see this effect, we consider the difference quantity $(Z_{4n})_{\phi=\pi}$ - $(Z_{4n})_{\text{min}}$ with respect to the loss parameter $|r|^2$ as shown in Fig. \ref{fig_5} (a). Here, for example, in case of $\Bar{N} = 3$ at $|r|^2=0$, $(Z_{4n})_{\text{min}}$ for SFCS $\approx 0.37$ and for ECSS $\approx 0.25$. While in Fig. \ref{fig_5} (b), we plot the peak height from the lowest point of the CS at $\phi=\pi$ with photon loss $|r|^2$. From Fig. \ref{fig_5} we can see that, at $\Bar{N} = 3$, SFCS shows loss robustness in comparison to the ECSS. However, in higher photon number, here $\Bar{N} = 100$, extra foldness rapidly degraded with loss. Negative values of peak height, in Fig. \ref{fig_5} (b), appear because the peak of SFCS goes lower than the lowest point of the CS at $\phi=\pi$.

In order to calculate the phase sensitivity, we use the standard error propagation formula \cite{Sharma:24}. In case of $Z_{4n}$-detection scheme, error propagation formula gives
\begin{equation}
 \Delta\phi_{Z_{4n}} = \frac{\sqrt{\langle \hat{Z}_{4n}\rangle-\langle \hat{Z}_{4n}\rangle^2}}{\left|\frac{\partial\langle \hat{Z}_{4n} \rangle}{\partial\phi}\right|},\label{z4n}
\end{equation}
where $\hat{Z}_{4n}^2=\hat{Z}_{4n}$ and $\left|\frac{\partial\langle \hat{Z}_{4n} \rangle}{\partial\phi}\right|$ is the variation in $\langle\hat{Z}_{4n}\rangle$ with phase $\phi$. From Eq. \eqref{z4nphase} we can write 
\begin{equation}
\begin{split}
\frac{\partial\langle \hat{Z}_{4n}\rangle}{\partial\phi} =|\mathcal{N}|^2e^{-|\alpha|^2}\frac{p^{4n}}{(4n)!}\Bigg( Xe^{-p+|\alpha|^2} \Big(\frac{4n}{p} - 1\Big) p'\\
- 2Vq' \sin(q) + 8 V \frac{n}{p} p' \cos(q)  + 2 Ye^{-q}\Big(\frac{4n}{p} p' - q'\Big)\Bigg),\label{eq z4nd}
\end{split} 
\end{equation}
where $p'= \frac{1}{2}|\alpha|^2|t|^2\sin^2\phi,~x'= -\frac{1}{2}|t|^2\sin^2\phi,~q' = -\frac{1}{2}|\alpha|^2|t|^2\sin^2\phi$. Using Eq. \eqref{z4nphase} and \eqref{eq z4nd} in Eq. \eqref{z4n}, we calculate the phase sensitivity for $Z_{4n}$-detection scheme. For $Z$-detection scheme, simply put $n = 0$ in the phase sensitivity of the $Z_{4n}$-detection case.

\begin{figure}
\includegraphics[width = \linewidth]{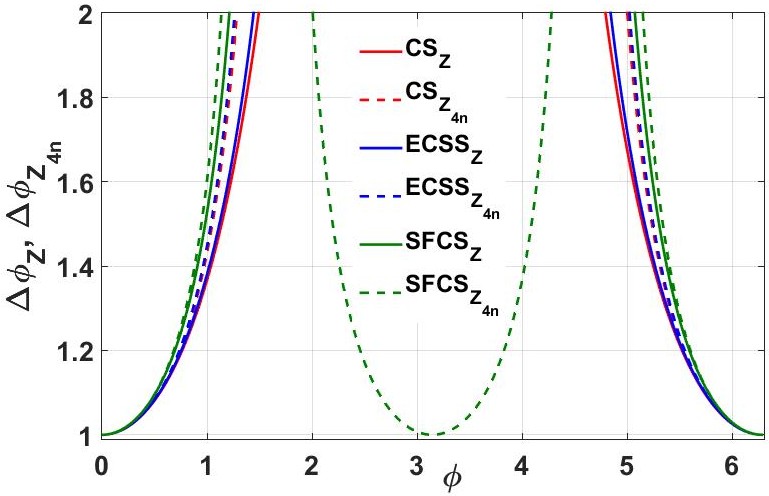}\caption{\label{fig_33} The Fig. shows the variation of $\Delta\phi$ with $\phi$ for all three states with  ${Z}$- detection and ${Z}_{4n}$-detection scheme for $\Bar{N} = 3$. }
\end{figure}

In order to compare the phase sensitivity of both detection schemes for different input states, we consider the ratio $\Delta\phi/SNL$ for the sake of equal energy inputs. In the case of higher photon numbers ($\Bar{N} = 100$), all the states show phase sensitivity similar to the coherent state case. However, in the low photon case ($\Bar{N} = 3$), we get some interesting results. Fig. \ref{fig_33} shows the $\Delta\phi/SNL$ with $\phi$ for both detection schemes for $\Bar{N} = 3$. From Fig. \ref{fig_33}, we draw the following conclusions:\\
(i) All the states saturate the SNL, as expected, because on one of the input ports, we have considered the vacuum state. Therefore, one can expect phase sensitivity below the SNL by using a coherent state instead of the vacuum state. (ii) In case of $Z_{4n}$-detection scheme, SFCS shows extra working points to achieve the SNL in comparison to the other states. 

The above results hold in the photon lossless case. While in lossy situation, all the states show similar degradation with photon loss in both detection schemes.

\textit{Conclusions:}\label{section 3}
In this letter, we have proposed a new detection scheme analogous to $Z$-detection, termed as ``$Z_{4n}$-detection". We have investigated the performance of $Z_{4n}$-detection and conventional $Z$-detection schemes for different input states, namely coherent state (CS), even coherent superposition state (ECSS), and superposition of four coherent states (SFCS). Our analysis shows that the $Z_{4n}$-detection scheme provides resolution enhanced with double foldness in comparison to the standard $Z$-detection for ECSS and SFCS, while no such improvement is observed for CS. Furthermore, we find that the resolution improves with increasing mean photon number, as indicated by the contraction in FWHM.

We have also examined the robustness of these schemes against photon loss. In the low photon regime, SFCS demonstrates comparatively better resilience than ECSS; however, at higher photon numbers, the advantage of additional foldness rapidly disintegrates under loss. The phase sensitivity analysis shows that all considered states saturate the SNL due to the presence of vacuum in one input port. While SFCS in the $Z_{4n}$-detection scheme offers additional working points to achieve optimal sensitivity.

In conclusion, our results highlight that although $Z_{4n}$-detection can enhance resolution and provide multiple operating points in the lossless regime, its practical advantage is significantly affected by photon loss, especially at higher photon numbers. These findings may be useful for optimizing detection strategies in quantum interferometric measurements using nonclassical states. Consequently, $Z_{4n}$-detection scheme holds promise as an alternative detection scheme for quantum imaging and quantum sensing applications.

\textit{Acknowledgments:}
PS acknowledges UGC for the UGC Research Fellowship. Authors would like to thank Dr. Gaurav Shukla for his useful comments and suggestions. DKM acknowledges financial supports from the ANRF(SERB), New Delhi for CRG Grant (CRG/2021/005917), I-Hub Quantum Technology Foundation, IISER Pune for the Chanakya Doctoral Fellowship grant (I-HUB/DF/2022-23/04) and Institution of Eminence (IoE), Banaras Hindu University, Varanasi, India for the Bridge Grant. MKM would like to thank Shri Nilesh M. Desai (Director, SAC, Ahmedabad) and Dr. Rashmi Sharma for their encouragement and support.

\appendix
\section{Calculation with SFCS and vacuum state as inputs}
The final state of the system at the output of the MZI
\begin{equation}
\begin{split}
     |\psi_{out}\rangle= \mathcal{N}\Big(|F_1\rangle
     +|F_2\rangle +|F_3\rangle
+|F_4\rangle\Big)_{a,b,E_a,E_b},
\end{split}
\end{equation}
where, $|\mathcal{N}|$ is given in Eq. \eqref{N0}, and $|F_1\rangle,~|F_2\rangle,~|F_3\rangle$ and $|F_4\rangle$ are given as
\begin{equation}
        \begin{split}
        |F_1\rangle=|\alpha\vartheta,\alpha\varsigma,{\Bar{r}\alpha e^{i\phi}},{i\Bar{r}\alpha}\rangle,
        |F_2\rangle=|\beta\vartheta,\beta\varsigma,{\Bar{r}\beta e^{i\phi}},{i\Bar{r}\beta}\rangle,\\
        |F_3\rangle=|\gamma \vartheta,\gamma\varsigma,{\Bar{r}\gamma e^{i\phi}},{i\Bar{r}\gamma}\rangle,
        |F_4\rangle=|\delta\vartheta,\delta\varsigma,{\Bar{r}\delta e^{i\phi}},{i\Bar{r}\delta}\rangle,\label{eqA3}
        \end{split}
\end{equation}
where $\vartheta=ite^{\frac{i\phi}{2}}\sin\left(\frac{\phi}{2}\right)$, $\varsigma=ite^{\frac{i\phi}{2}}\cos\left(\frac{\phi}{2}\right)$ and $\Bar{r}=ir/\sqrt{2}$.
The density operator ($\hat{\rho}$) for the final state of the system is written as
\begin{equation}
\begin{split}
\hat{\rho}= |\psi\rangle_{out~out}\langle\psi|=|N|^2\left(|F_1\rangle\langle F_1|\right.\\
+|F_2\rangle\langle F_2|+|F_3\rangle\langle F_3|+|F_4\rangle\langle F_4|\\
+|F_1\rangle\langle F_2|+|F_2\rangle\langle F_1|+|F_1\rangle\langle F_3|\\
+|F_3\rangle\langle F_1|+|F_1\rangle\langle F_4|+|F_4\rangle\langle F_1|\\
+|F_2\rangle\langle F_3|+|F_3\rangle\langle F_2|+|F_2\rangle\langle F_4|\\
+\left.|F_4\rangle\langle F_2|+|F_3\rangle\langle F_4|+|F_4\rangle\langle F_3|\right). \label{221abc}
\end{split}
\end{equation}
.

\bibliography{apssamp}

\end{document}